\documentclass{iopart}

\usepackage{citesort}
\usepackage{graphicx}

\usepackage{iopams,setstack}
\usepackage{epsf}
\usepackage{cite}
\usepackage{enumerate}

\newcommand{\bd}{\begin{displaymath}}
\newcommand{\ed}{\end{displaymath}}
\newcommand{\be}{\begin{equation}}
\newcommand{\ee}{\end{equation}}
\newcommand{\ba}{\begin{eqnarray}}
\newcommand{\ea}{\end{eqnarray}}

\begin{document}

\paper[Trajectory-based interpretation for photons and massive
particles]{Trajectory-based interpretation of Young's experiment,
the Arago-Fresnel laws and the Poisson-Arago spot for photons and
massive particles}

\author{Milena Davidovi\'c$^1$, \'Angel S. Sanz$^2$,
Mirjana Bo\v zi\'c$^3$, Du\v san Arsenovi\'c$^3$ and Dragan Dimi\'c$^4$}

\address{$^1$Faculty of Civil Engineering, University of Belgrade,
Bulevar Kralja Aleksandra 73, 11000 Belgrade, Serbia}

\address{$^2$Instituto de F\'{\i}sica Fundamental (IFF-CSIC),
Serrano 123, 28006 Madrid, Spain}

\address{$^3$Institute of Physics, University of Belgrade,
Pregrevica 118, 11080 Belgrade, Serbia}

\address{$^4$4Faculty of Natural and Mathematical Sciences,
University of Ni\v s, Ni\v s, Serbia}

\eads{\mailto{davidovic.milena@gmail.com},\mailto{asanz@iff.csic.es},
\mailto{bozic@ipb.ac.rs},\mailto{arsenovic@ipb.ac.rs},
\mailto{divladq@hotmail.com}}

\begin{abstract}
We present a trajectory based interpretation for Young's experiment,
the Arago-Fresnel laws and the Poisson-Arago spot. This approach is
based on the equation of the trajectory associated with the quantum
probability current density in the case of massive particles, and
the Poynting vector for the electromagnetic field in the case of
photons. Both the form and properties of the evaluated photon
trajectories are in good agreement with the averaged trajectories of
single photons observed recently in Young's experiment by
Steinberg's group at the University of Toronto. In the case of the
Arago-Fresnel laws for polarized light, the trajectory
interpretation presented here differs from those interpretations
based on the concept of ``which-way'' (or ``which-slit'')
information and quantum erasure. More specifically, the observer's
information about the slit that photons went through is not relevant
to the existence of interference; what is relevant is the form of
the electromagnetic energy density and its evolution, which will
model consequently the distribution of trajectories and their
topology. Finally, we also show that the distributions of end points
of a large number of evaluated photon trajectories are in agreement
with the distributions measured at the screen behind a circular
disc, clearly giving rise to the Poisson-Arago spot.
\end{abstract}


%03.65.Ta %Foundations of quantum mechanics; measurement theory
%03.75.Dg %Atom and neutron interferometry
%42.50.-p %Quantum optics
%42.50.Xa %Optical tests of quantum theory
%37.25.+k
%42.25.Hz %Interference

%PACS
%03.65.-w : Quantum mechanics (general)
%03.65.Ta : Foundations of quantum mechanics; measurement theory
%03.65.Xp : Tunneling, traversal time, quantum Zeno dynamics
%07.79.Cz : Scanning tunneling microscopes
%68.37.Ef : Scanning tunneling microscopy (including chemistry induced with STM)
%73.23.Hk : Coulomb blockade; single-electron tunneling
%82.20.Xr : Quantum effects in rate constants (tunneling, resonances, etc.)
%03.75.-b : Matter waves
%42.25.Hz : Interference

%\submitto{\PS}

%\maketitle

%%%%%%%%%%%%%%%%%%%%%%%%%%%%%%%%%%%%%%%%%%%%%%%%%%%%%%%%%%%%%%%%%%%%%%%
%%%%%%%%%%%%%%%%%%%%%%%%%%%%%%%%%%%%%%%%%%%%%%%%%%%%%%%%%%%%%%%%%%%%%%%

\section{Introduction}
 \label{sec1}

Very refined and ingenious interferometers and detectors for
electrons \cite{pozzi:AJP:1976}, neutrons
\cite{rauch:PLA:1974,rauch-bk,klein:EuphysNews:2009}, atoms
\cite{pritchard:PRL:1991,mlynek:Nature:1997,berman-bk}, molecules
\cite{berman-bk,arndt:FoundPhys:2012} and photons
%\cite{parker:AJP:1971,parker:AJP:1972,weis:AJP:2008}
\cite{parker:AJP:1971,weis:AJP:2008} have been devised to
demonstrate that the quantum interference pattern can be build up by
means of the accumulation of single detection events. Even before
the realization of these experiments, De Broglie
%\cite{broglie-bk-fr,broglie-bk-en}
\cite{broglie-bk-fr} and Bohm
%\cite{bohm:PR:1952-1,bohm:PR:1952-2}
\cite{bohm:PR:1952-1} argued that particles with mass possess
simultaneously wave and particle properties, and would move within
an interferometer along trajectories determined by the guidance
equation
\be
 {\bf v} = \frac{d{\bf r}}{dt} = \frac{\nabla S({\bf r},t)}{m} ,
 \label{eq1}
\ee
where $S({\bf r},t)$ is the phase of the particle wave function
\be
 \Psi({\bf r},t) = |\Psi({\bf r},t)| e^{iS({\bf r},t)/\hbar}
 \label{eq2}
\ee
which satisfies the time-dependent Schrödinger equation.

Using the method proposed by De Broglie and Bohm, Philipidis \etal
\cite{dewdney:NuovoCim:1979} plotted the trajectories of massive
particles in the double slit experiment
\cite{dewdney:NuovoCim:1979}, Dewdney showed trajectories for
neutrons inside a (neutron) interferometer \cite{dewdney:PLA:1985},
and Sanz and Miret-Artés explained the Talbot effect for atoms by
plotting their associated trajectories behind a diffraction grating
\cite{sanz:JCP-Talbot:2007}.

Motivated by the works of Laukien \cite{laukien:Optik:1952},
presented in \cite{bornwolf-bk}, and Prosser
%\cite{prosser:ijtp:1976-1,prosser:ijtp:1976-2},
\cite{prosser:ijtp:1976-1}, Davidovi\'c \etal explained
\cite{sanz:PhysScrPhoton:2009} the emergence of interference
patterns in experiments with photons by determining electromagnetic
energy (EME) flow lines behind an interference grating. The equation
of such EME flow lines reads as
\be
 \frac{d{\bf r}}{ds} = \frac{{\bf S}({\bf r})}{cU({\bf r})} ,
 \label{eq3}
\ee
where $s$ denotes a certain arc-length along the corresponding path,
${\bf S}({\bf r})$ is the real part of the complex-valued Poynting
vector,
\be
 {\bf S}({\bf r}) = \frac{1}{2}\ \!
  {\rm Re} \left[{\bf E}({\bf r}) \times {\bf H}^*({\bf r})\right] ,
 \label{eq4}
\ee
and $U({\bf r})$ is the time-averaged EME density,
\be
 U({\bf r}) = \frac{1}{4}\ \! \left[
  \epsilon_0 {\bf E}({\bf r}) \cdot {\bf E}^*({\bf r})
  + \mu_0 {\bf H}({\bf r}) \cdot {\bf H}^*({\bf r}) \right] .
 \label{eq5}
\ee
Here ${\bf E}({\bf r})$ and ${\bf H}({\bf r})$ are respectively the
spatial part of the electric and
magnetic field vectors, which satisfy Maxwell's equations and have
been assumed to be harmonic, i.e.,
\be
 \begin{array}{c}
  \tilde{\bf E}({\bf r}) = {\bf E}({\bf r}) e^{-i\omega t} , \\
  \tilde{\bf H}({\bf r}) = {\bf H}({\bf r}) e^{-i\omega t} .
 \end{array}
 \label{eq6}
\ee
Davidovi\'c \etal pointed out \cite{sanz:PhysScrPhoton:2009} that it
is useful to write the equation of the Bohmian trajectories for massive
particles (\ref{eq1}) in terms of the probability current density,
\be
 {\bf J}({\bf r},t) = \frac{\hbar}{2im}
  \left[ \Psi \nabla \Psi^* - \Psi^* \nabla \Psi \right] ,
 \label{eq7}
\ee
because from the latter form one can recast the guidance equation
(\ref{eq1}) as
\be
 \frac{d{\bf r}}{dt} = \frac{{\bf J}({\bf r},t)}
  {|\Psi({\bf r},t)|^2} ,
 \label{eq8}
\ee
from which one may conclude that the equation for the  EME flow
lines and the equation of the Bohmian trajectories for massive
particles have the same form. In other words, the Poynting vector in
the case of photons plays the same role as the quantum probability
current density in the case of particles with a mass.

This analogy is even more apparent in cases where the spatial parts
of the magnetic and electric fields can be expressed in terms of a
scalar function that satisfies the Helmholtz equation
\cite{sanz:PhysScrPhoton:2009,sanz:AnnPhysPhoton:2010}.

M. Gondran and A. Gondran \cite{gondran:AJP:2010} explained the
appearance of the Poisson-Arago spot behind an illuminated circular
disc using EME flow lines. In this way, they showed how such flow
lines answer the question about diffraction phenomena presented two
centuries ago by the French Academy ``{\it deduce by mathematical
induction, the movements of the rays during their crossing near the
bodies.''}

Recently, average trajectories of single photons in a double slit
experiment were observed experimentally for the first time by Kocis
\etal \cite{kocsis:Science:2011}. Their result motivated us to
apply the method of EME flow lines to numerically evaluate photon
trajectories behind the double-slit grating with the same parameters
as in the Kocis \etal experiment. We show these results in
Section~\ref{sec2}. In Section~\ref{sec3} we study how polarizers
put behind the slits
affect the photon trajectories, thus providing a trajectory
interpretation for the Arago-Fresnel laws. By adding orthogonal
polarizers behind the slits, Kocis \etal could observe average
photon trajectories in the presence of orthogonal polarizers and
check directly this interpretation. Section~\ref{sec4} is devoted
to the trajectory based interpretation of the Poisson-Arago spot.

%%%%%%%%%%%%%%%%%%%%%%%%%%%%%%%%%%%%%%%%%%%%%%%%%%%%%%%%%%%%%%%%%%%%%%%
%%%%%%%%%%%%%%%%%%%%%%%%%%%%%%%%%%%%%%%%%%%%%%%%%%%%%%%%%%%%%%%%%%%%%%%

\section{EME flow lines - average photon trajectories in Young's
interferometer}
\label{sec2}

Let us consider a monochromatic electromagnetic wave in vacuum
incident onto a two slit grating located on the $XY$ plane, at $z=0$.
In order to simplify the treatment we will assume that the electric and
magnetic fields do not depend on the $y$ coordinate. This assumption
is justified when the slits are parallel to the $y$ axis and their
width along the $y$ axis is much larger than the width along the
$x$ axis.
In such a case from Maxwell's equations one obtains two independent
sets of equations: one involving the $H_x$ and $H_z$ components of the
magnetic field and the $E_y$ component of the electric field (commonly
referred as $E$-polarization), and another involving $E_x$, $E_z$ and
$H_y$ ($H$-polarization). As shown in \cite{sanz:PhysScrPhoton:2009},
the electric and magnetic fields behind the grating are given by
\ba
 {\bf E}({\bf r}) & = &
   -\frac{i\beta}{k}\frac{\partial \Psi}{\partial z}\ \! {\bf e}_x
   +\frac{i\beta}{k}\frac{\partial \Psi}{\partial x}\ \! {\bf e}_z
   +\alpha \Psi {\bf e}_y ,
 \label{eq9} \\
 {\bf H}({\bf r}) & = &
    \frac{i\alpha}{\omega\mu_0}\frac{\partial \Psi}{\partial z}\ \!
     {\bf e}_x
   -\frac{i\alpha}{\omega\mu_0}\frac{\partial \Psi}{\partial x}\ \!
     {\bf e}_z
   +\frac{k\beta e^{i\varphi}}{\omega\mu_0}\ \! \Psi {\bf e}_y ,
 \label{eq10}
\ea
where $\Psi$ is a scalar function that satisfies the Helmholtz equation
and the boundary conditions at the grating. The solution for $\Psi$ can
be written as a Fresnel-Kirchhoff integral,
\be
 \Psi(x,z) = \sqrt{\frac{k}{2\pi z}}\ \! e^{ikz - i\pi/4}
  \int_{-\infty}^\infty \psi(x',0^+) e^{ik(x-x')^2/2z} dx' ,
 \label{eq11}
\ee
where $\psi(x',0^+)$ is the wave function just behind the grating.

\begin{figure}
 \begin{center}
 \epsfxsize=6.25cm {\epsfbox{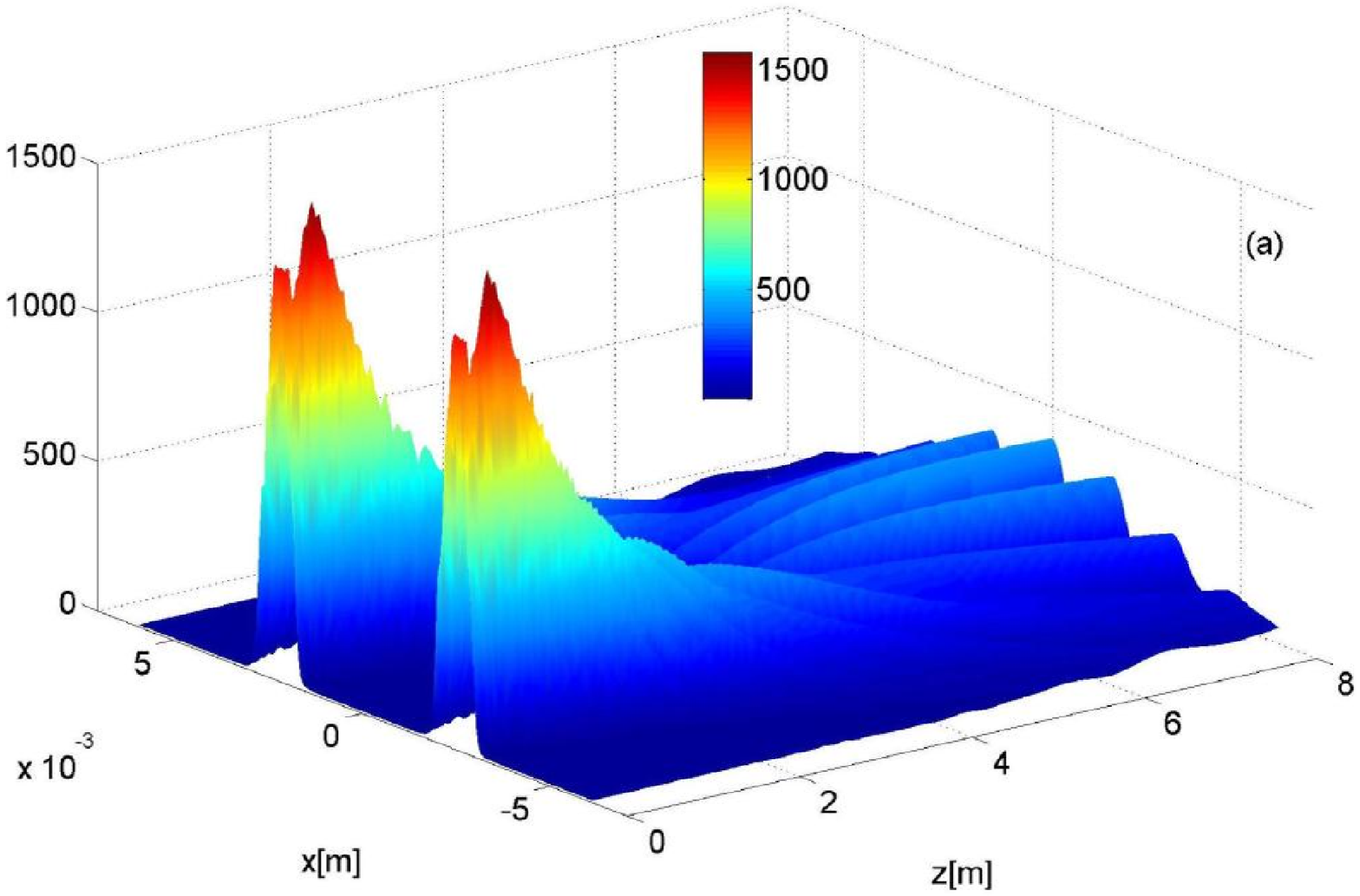}}
 \epsfxsize=6.25cm {\epsfbox{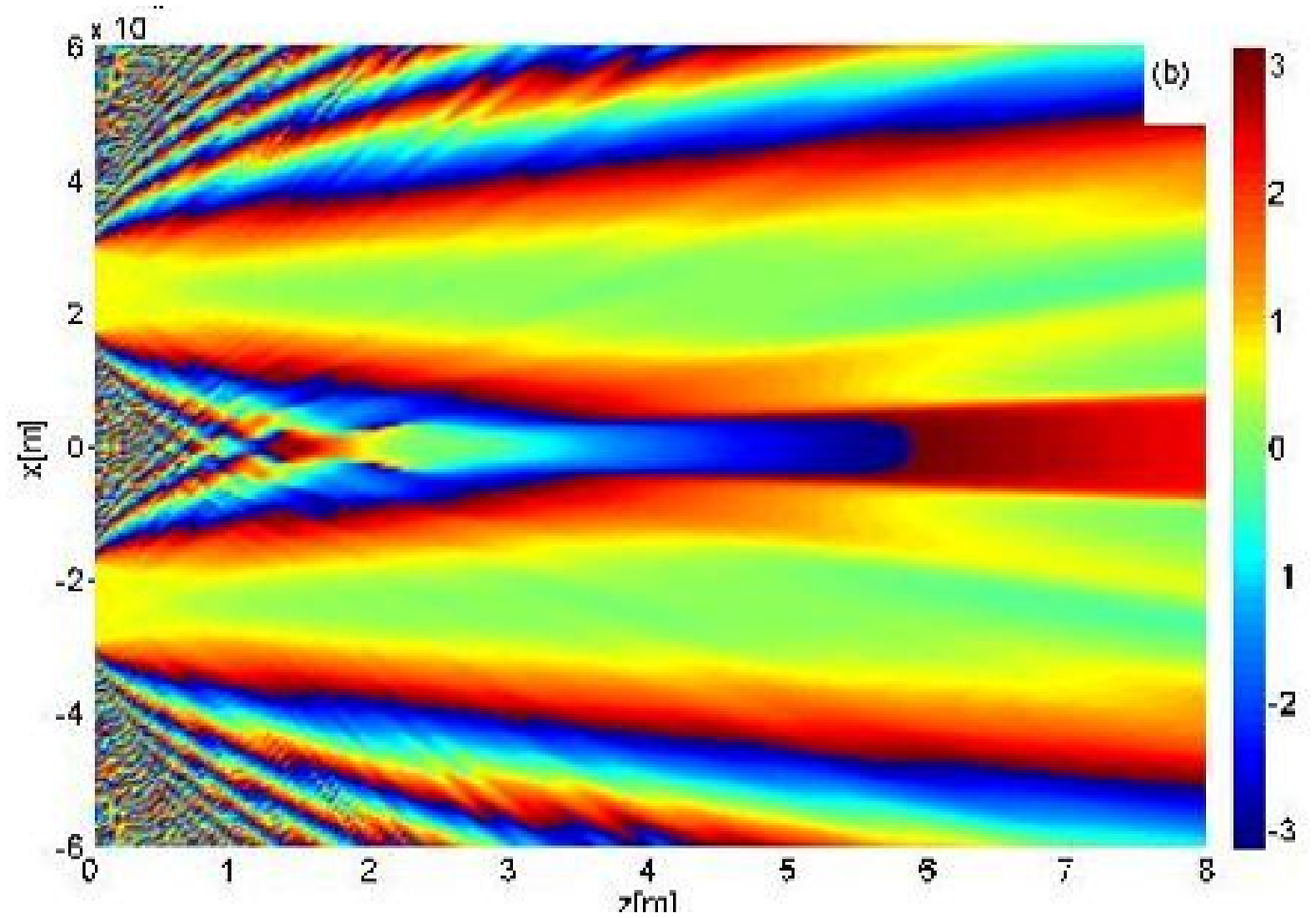}}
 \epsfxsize=6.25cm {\epsfbox{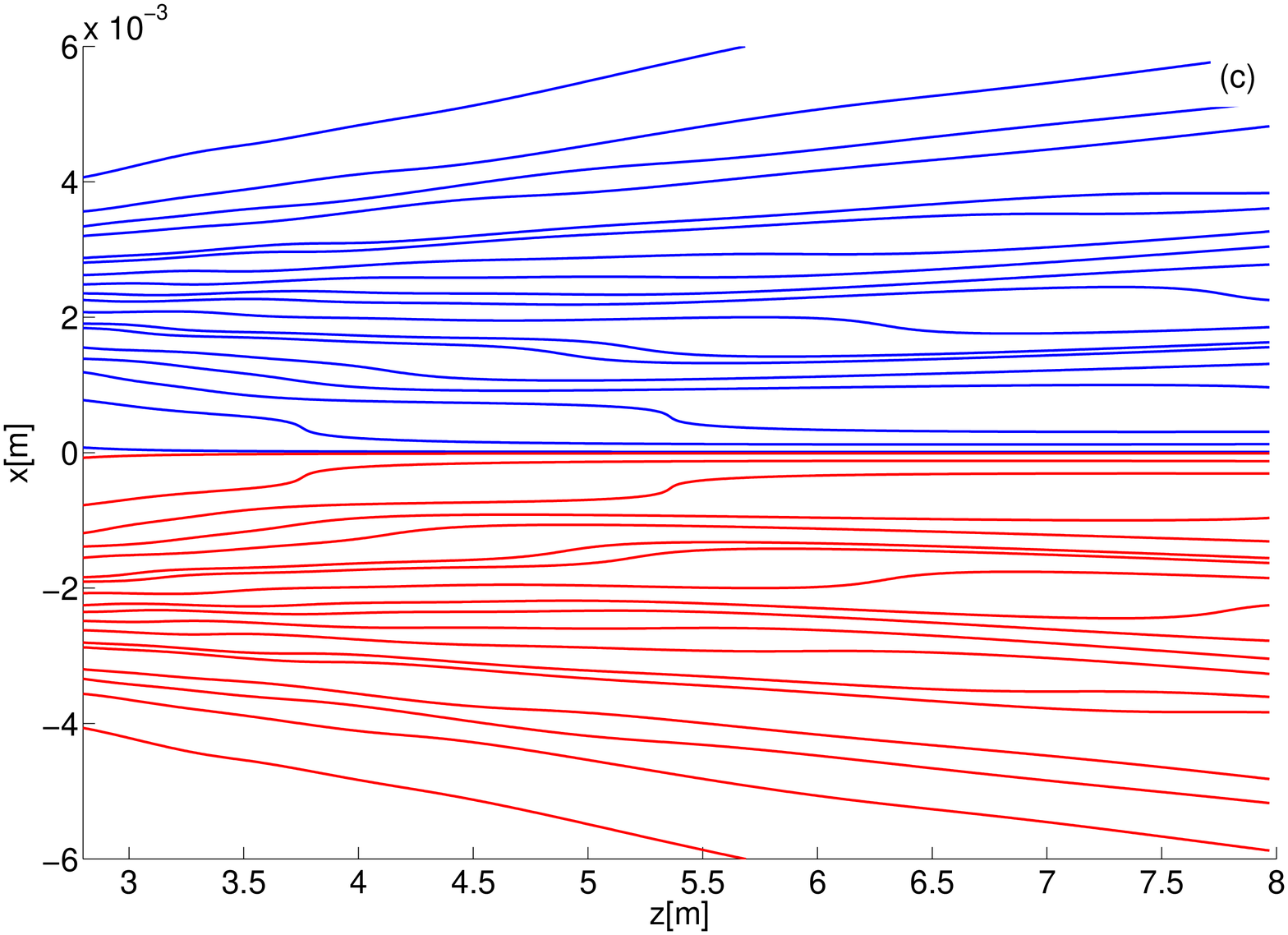}}
 \caption{\label{fig1}
  EME density (a), phase of $\Psi$ (b) and photon trajectories (c)
  behind a two-slit Gaussian grating. The parameters are chosen as in
  the experiment carried out by Kocsis \etal:
  $\sigma_1 = \sigma_2 = 0.3$~mm, $\mu_1 = \mu_2 = 2.35$~mm,
  $a_1 = a_2 = 1.8\sigma_1$ and $\lambda = 943$~nm.
  The initial polarization is lineal.
  The initial $x$-coordinates for the trajectories are calculated from
  (\ref{eq16}), where $u$ takes 19 equidistant values within the
  interval $[0.02,0.98]$.}
 \end{center}
\end{figure}

We consider a grating with two Gaussian slits \cite{feynman-bk-1a},
so that the wave function just behind the grating is given by
\be
 \psi(x',0^+) = \psi_1(x',0^+) + \psi_2(x',0^+) ,
 \label{eq12}
\ee
where
\ba
 \psi_1(x',0^+) & = & \left(\frac{1}{2\pi\sigma_1^2}\right)^{1/4}
  e^{-(x'-\mu_1)^2/4\sigma_1^2} w(x'-\mu_1,a_1) ,
 \label{eq13} \\
 \psi_2(x',0^+) & = & \left(\frac{1}{2\pi\sigma_2^2}\right)^{1/4}
  e^{-(x'-\mu_2)^2/4\sigma_2^2} w(x'-\mu_2,a_2) ,
 \label{eq14}
\ea
and $w(x,a)$ is the window function,
\be
 w(x,a) = \left\{ \begin{array}{cc}
  1, & x \in [-a,a] \\
  0, & x \notin [-a,a] \end{array} \right. .
 \label{eq15}
\ee
The EME flow lines (i.e., the average photon trajectories from the
experiment carried out by Kocsis \etal) are obtained from equations
(\ref{eq3})-(\ref{eq5}) and (\ref{eq9})-(\ref{eq15}).
In figure~\ref{fig1}, 19 photon trajectories per slit are shown.
The initial x coordinates of the flow lines are chosen to be
\be
 x_s = \mu_i + \sigma_i F^{-1} (u) ,
 \label{eq16}
\ee
where $i \in [1,2]$ is the cardinal number of the slit and $F^{-1}(u)$
is the inverse of the Gaussian cumulative distribution function.
If the variable $u$ follows a uniform distribution, then the variable
$x_s$ will have a Gaussian distribution with mean value $\mu_i$ and
variance $\sigma_i$.

\begin{figure}
 \begin{center}
 \epsfxsize=6.25cm {\epsfbox{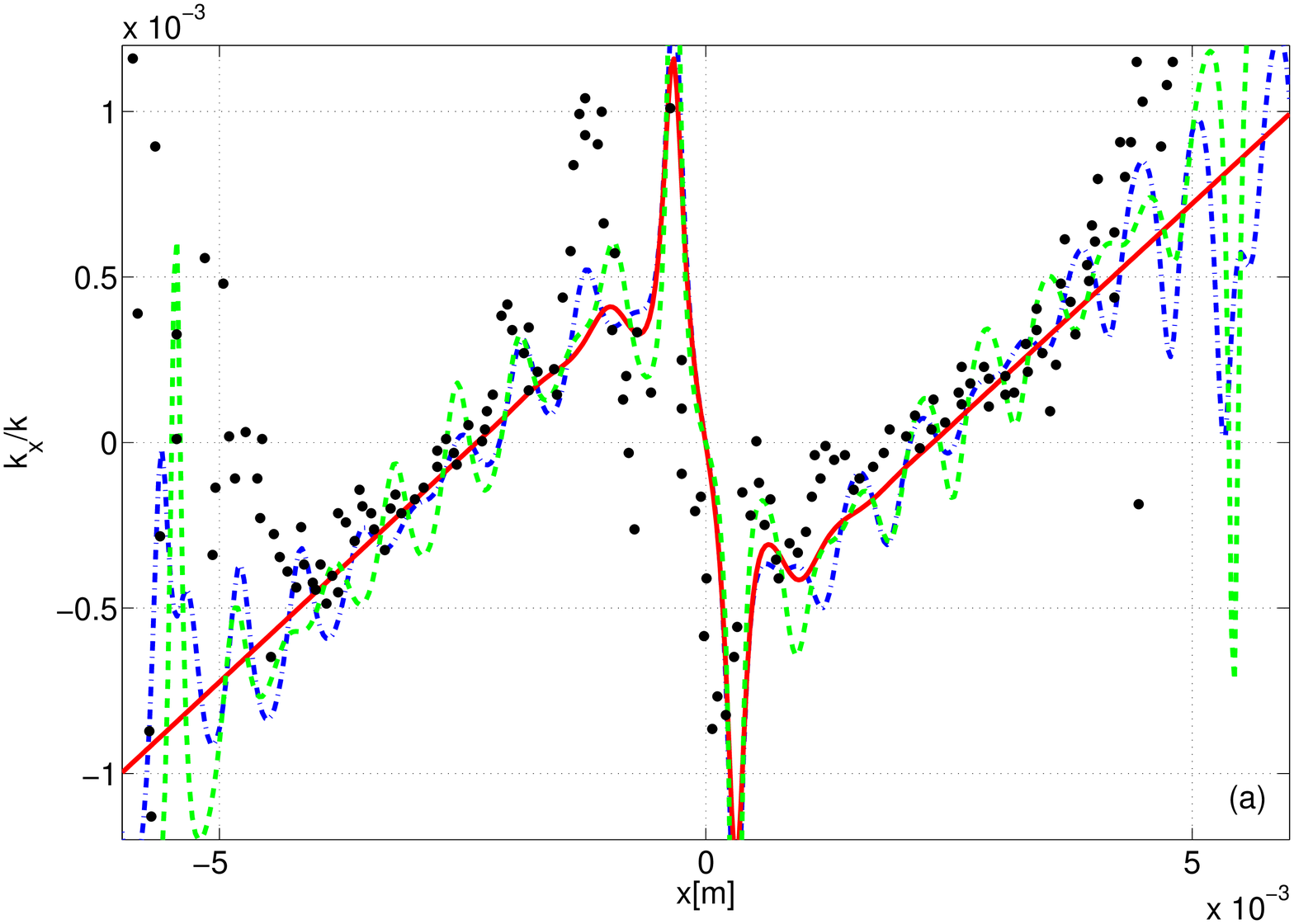}}
 \epsfxsize=6.25cm {\epsfbox{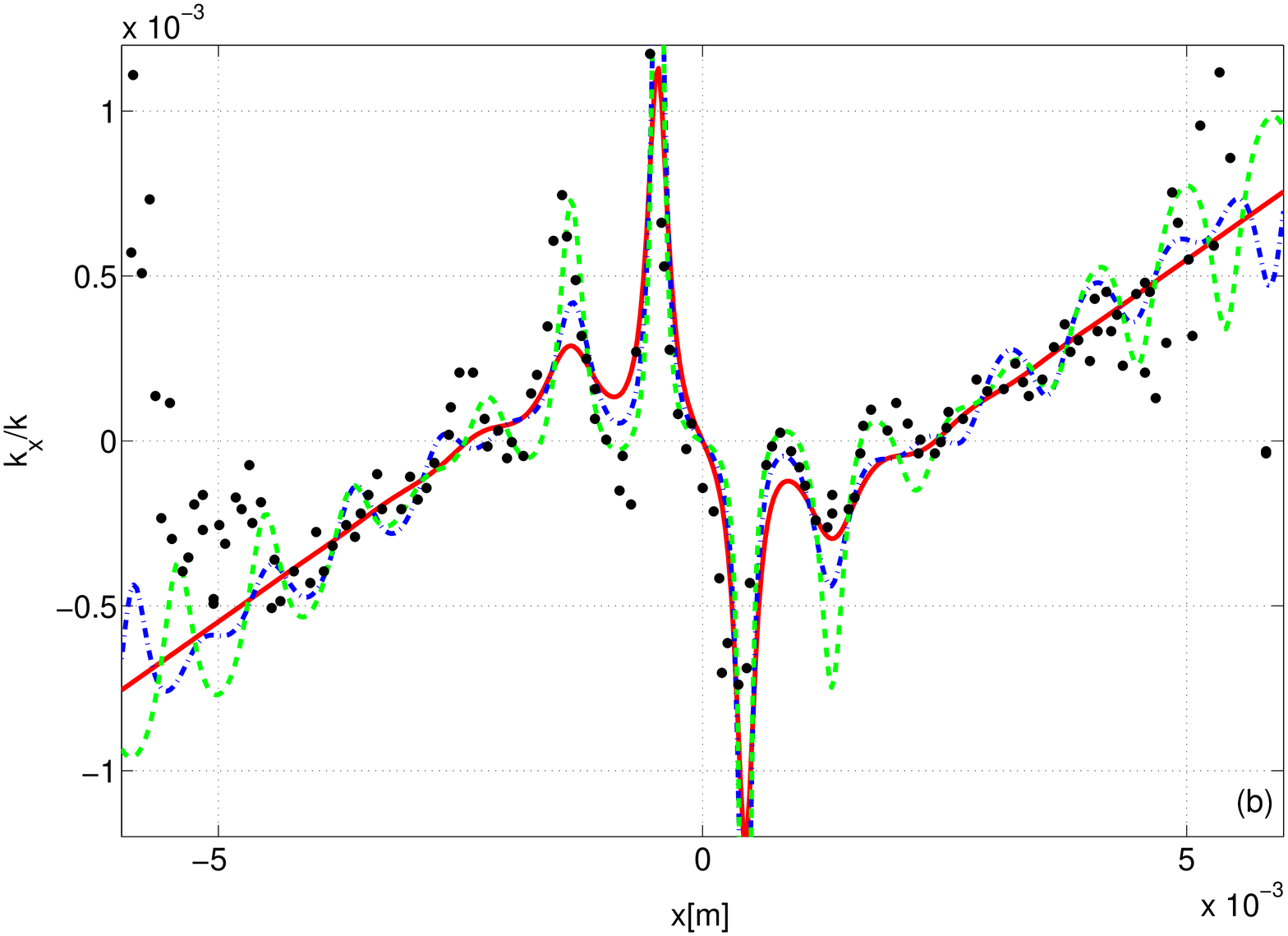}}
 \epsfxsize=6.25cm {\epsfbox{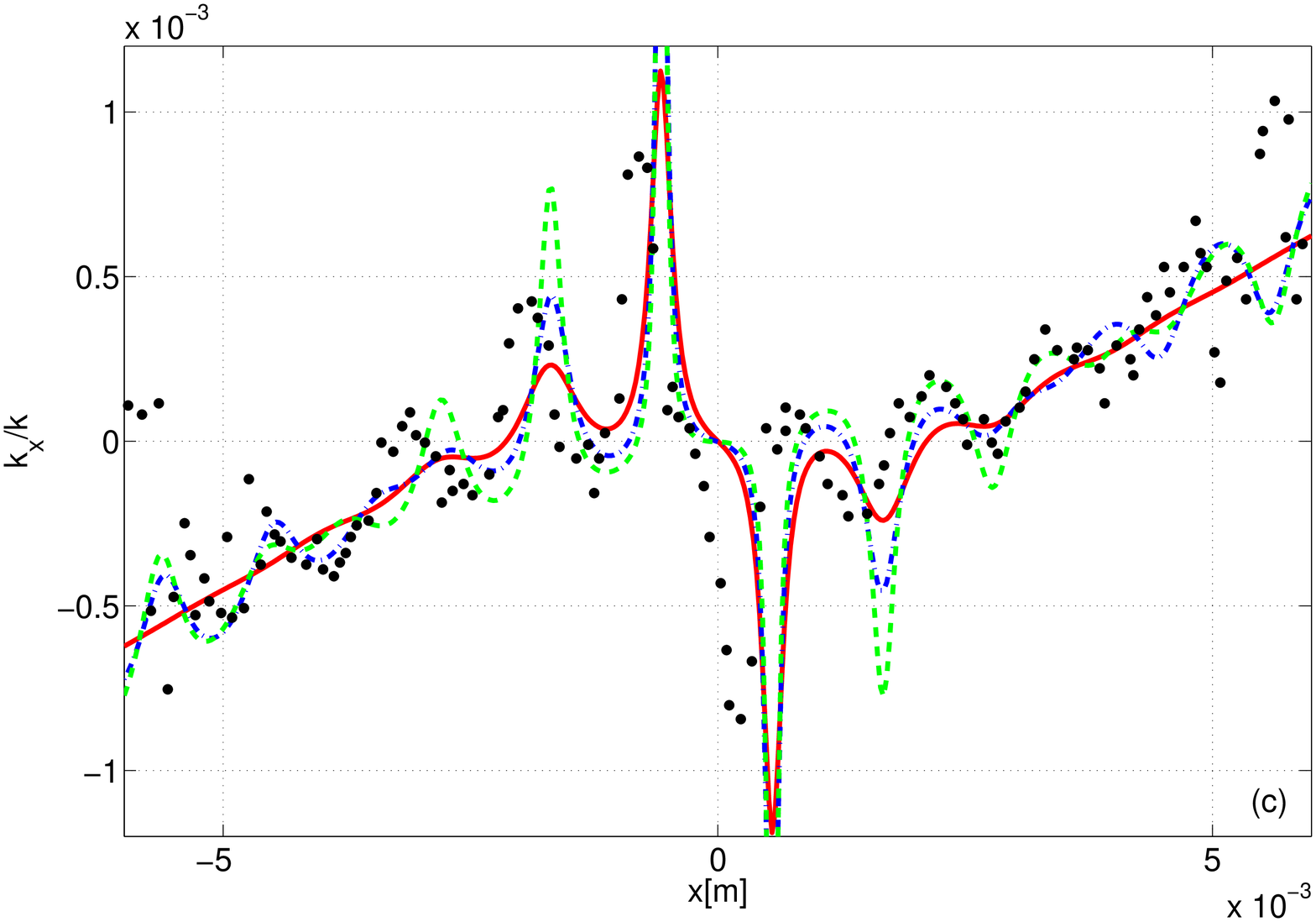}}
 \epsfxsize=6.25cm {\epsfbox{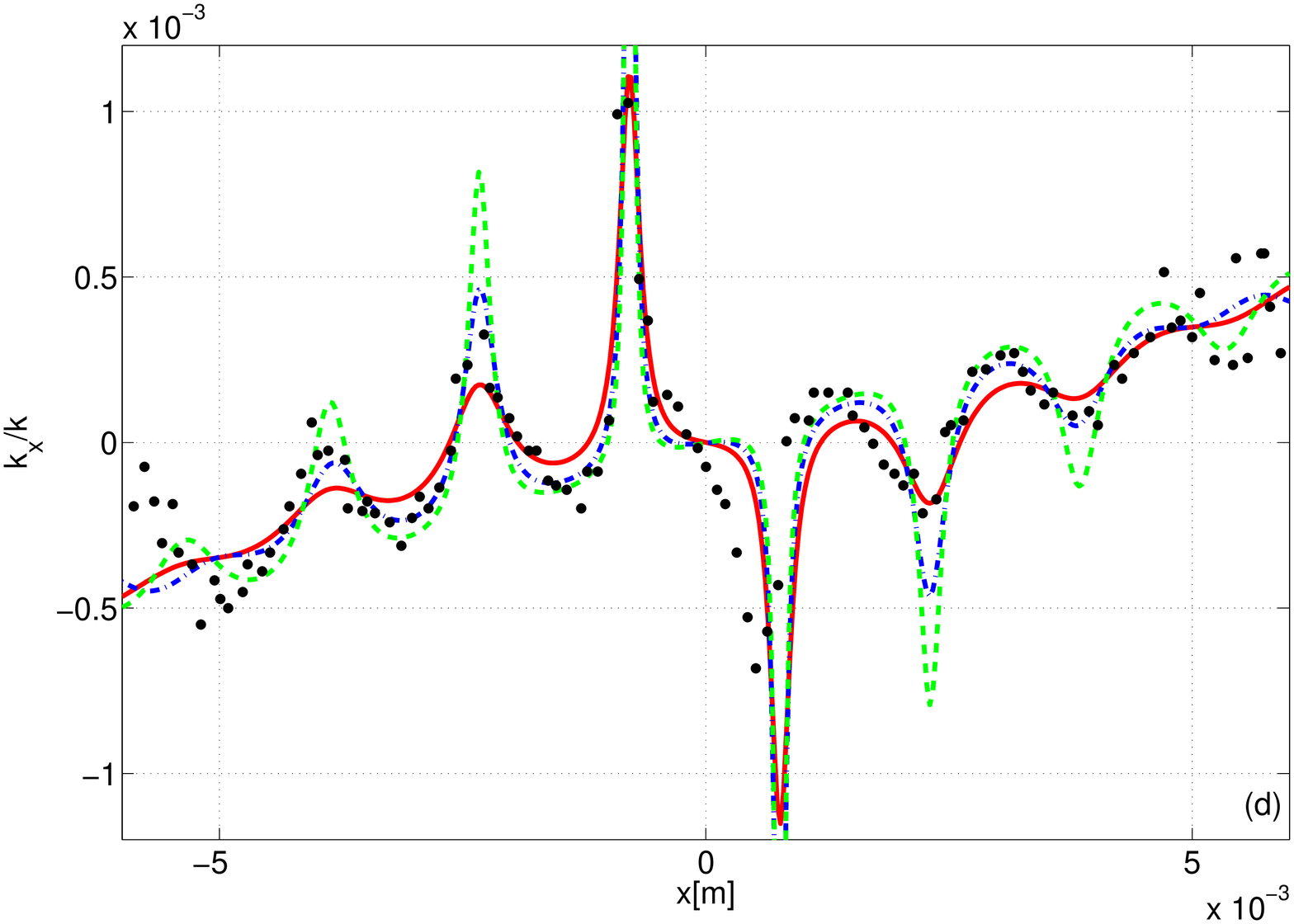}}
 \caption{\label{fig2}
  Transverse momentum along the transverse coordinate computed at four
  distances $z$ from the two slits: (a) $z = 3.2$~m, (b) $z = 4.5$~m,
  (c) $z = 5.6$~m and (d) $z = 7.7$~m. The red solid line denotes the
  calculation with full Gaussians, while the blue and green lines
  refer to calculations where the outgoing beams were truncated
  Gaussians with $a = 1.9\sigma$ and $a = 1.5\sigma$, respectively.
  To compare with, the experimental data (black circles) are also
  displayed. The parameters used for calculation are:
  $\sigma_1 = 0.307$~mm, $\sigma_2 = 0.301$~mm, $\mu_1 = 2.335$~mm,
  $\mu_2 = -2.355$~mm, $a_1 = 1.5\sigma_1$, $a_2 = 1.5\sigma_2$ and
  $\lambda = 943$~nm.}
 \end{center}
\end{figure}

By comparing the photon trajectories displayed in figure 1(c) with
those experimentally inferred, shown in figure 3 in
\cite{kocsis:Science:2011}, we notice a very good agreement. The
experimental average photon paths were reconstructed after
performing a weak measurement on the momentum of an ensemble of
photons and then a subsequent strong measurement of their position.
In figure~\ref{fig2}, we compare the experimental data coming from the
measurement of the relative weak transverse momentum values, $k_x/k$,
as a function of the transverse coordinate at four different distances
from the grating, with our theoretical curves obtained for three
different window functions.

Since in our case the light propagates in vacuum, the Poynting
vector can be identified with the density of electromagnetic
momentum \cite{barnett:PhilTRSA:2010}, so we have
\be
 \frac{k_x}{k} = \frac{S_x}{S} .
 \label{eq17}
\ee
The Fresnel-Kirchhoff integral \cite{bornwolf-bk} can be integrated
analytically for full Gaussians (for which the parameter of the
window function $a \to \infty$), while the integration has to be done
numerically for truncated Gaussians, as we did here.

%%%%%%%%%%%%%%%%%%%%%%%%%%%%%%%%%%%%%%%%%%%%%%%%%%%%%%%%%%%%%%%%%%%%%%%
%%%%%%%%%%%%%%%%%%%%%%%%%%%%%%%%%%%%%%%%%%%%%%%%%%%%%%%%%%%%%%%%%%%%%%%

\section{EME flow-line interpretation of the Arago-Fresnel laws}
\label{sec3}

According to a generalized version of the Arago-Fresnel laws, two
beams with the same polarization state interfere with each other
just as natural rays do, but no interference pattern will be
observable if the two interfering beams are linearly polarized in
orthogonal directions or elliptically polarized, with opposite
handedness and mutually orthogonal major axes. The most direct way
to verify these laws is by inserting mutually orthogonal polarizers
behind the slits of a double-slit grating.

The standard interpretation given to the disappearance of the
interference fringes after inserting mutually orthogonal polarizers
behind the slits is usually based on the Copenhagen notion of the
external observer's knowledge (information) about the path followed
by the photon, i.e., the slit traversed by the photon in its way to
the detection screen. Sanz et al \cite{sanz:AnnPhysPhoton:2010} and
Bo\v zi\'c \etal \cite{sanz:JRLR:2010} challenged this
interpretation by explaining the first and second Arago-Fresnel laws
considering EME flow lines behind the grating both in the presence
and in the absence of polarizers. In both cases EME flow lines
starting from slit 1 will end up in the side in front of slit 1,
while those starting from slit 2 will end up in the side in front of
slit 2, as also shown in quantum mechanics for matter particles
\cite{sanz-miret-1-bk}. However, the distribution of these EME flow
lines is different in each case. In the absence of polarizers, the
distribution shows interference fringes (figure 1(c) and figure 5 in
\cite{sanz:JRLR:2010}); in the presence of polarizers, the fringes
are absent (figure 3 and figure 6 in \cite{sanz:JRLR:2010}).

\begin{figure}
 \begin{center}
 \epsfxsize=7cm {\epsfbox{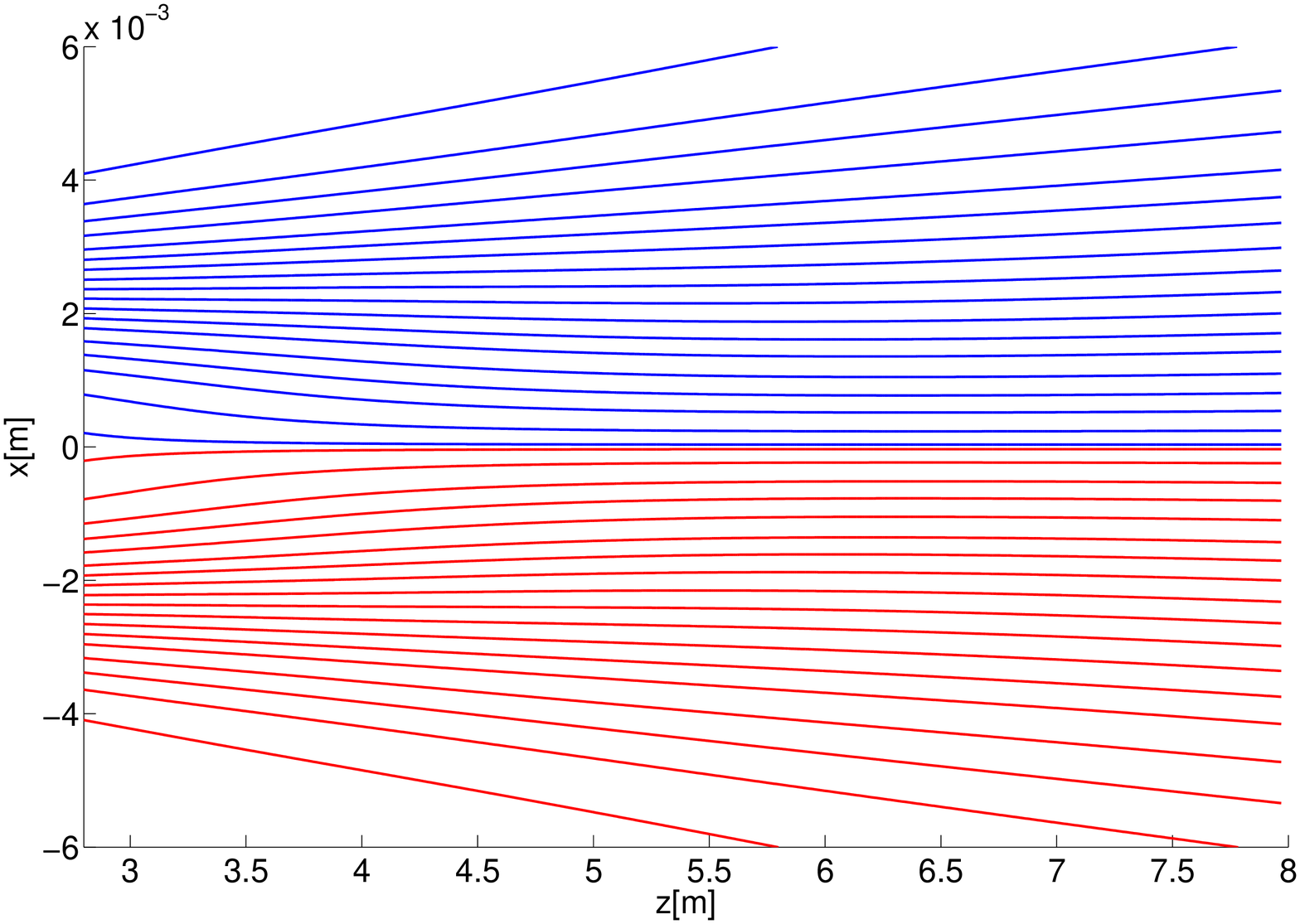}}
 \caption{\label{fig3}
  Photon trajectories behind a Gaussian double-slit grating followed
  by orthogonal polarizers. The parameters are the same as in
  figure~\ref{fig1}.}
 \end{center}
\end{figure}

As seen above, the average photon trajectories observed by Kocis et
al. \cite{kocsis:Science:2011} in the absence of polarizers agree
with our EME flow lines ---the photon paths presented in
figure~\ref{fig1}. In
order to verify the interpretation of the Arago-Fresnel laws based
on the EME flow lines, it would be interesting as well as
challenging to experimentally determine the average photon paths in
slightly modified experimental setup, by adding orthogonal
polarizers behind the slits. In such a case, we expect that the
corresponding experimentally inferred photon paths would look like
the trajectories presented in figure~\ref{fig3}.

%%%%%%%%%%%%%%%%%%%%%%%%%%%%%%%%%%%%%%%%%%%%%%%%%%%%%%%%%%%%%%%%%%%%%%%
%%%%%%%%%%%%%%%%%%%%%%%%%%%%%%%%%%%%%%%%%%%%%%%%%%%%%%%%%%%%%%%%%%%%%%%

\section{EME flow-line interpretation of the Poisson-Arago spot}
\label{sec4}

\begin{figure}
 \begin{center}
 \epsfxsize=7cm {\epsfbox{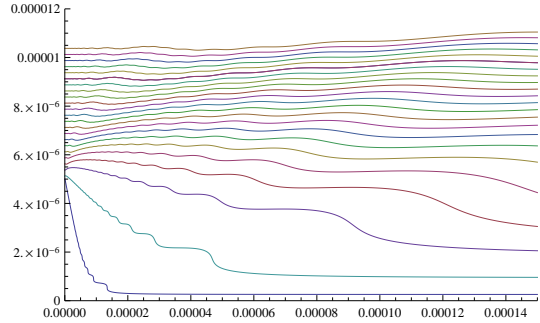}}
 \caption{\label{fig4}
  Photon trajectories in the $XZ$ plane behind a circular disc in the
  $XY$ plane, centered at $x = y = z = 0$ and with radius
  $R = 5$~$\mu$m. The disc is illuminated by a monochromatic light with
  wavelength $\lambda = 500$~nm. Because of the cylindrical
  symmetry of the problem, only trajectories having positive initial
  $x$ coordinate are presented.}
 \end{center}
\end{figure}

\begin{figure}
 \begin{center}
 \epsfxsize=7cm {\epsfbox{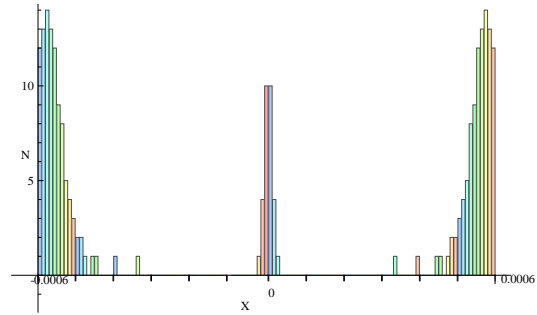}}
 \caption{\label{fig5}
  Histogram of the end points of photon trajectories at the distance
  $z = 15$~mm behind a circular disc, having radius $R = 0.5$~mm,
  illuminated by the monochromatic light with wavelength
  $\lambda = 500$~nm.}
 \end{center}
\end{figure}

It is well known that the experimental observation of the so-called
Poisson-Arago spot\footnote{This phenomenon is commonly regarded
simply as Poisson spot. However, we have added the name of Arago in
order to give him scientific credit, for it was him who provided the
experimental evidence for this light phenomenon.} by Arago led to
the acceptance of Fresnel's wave theory of light and the refutation
of Newton's corpuscular theory of light. Now, by numerically
evaluating EME flow lines behind a circular opaque disc, M. Gondran
and A. Gondran found that these lines can reach the bright
Poisson-Arago spot that appears at the center of the shadow region
generated by such a disc \cite{gondran:AJP:2010}. These authors then
argued that for a monochromatic wave in vacuum, the EME flow lines
correspond to the diffracted rays of Newton's Opticks, thus
concluding that after all Fresnel's wave theory may not be in
contradiction with the corpuscular interpretation. This result also
follows from our evaluation of EME flow lines (figure~\ref{fig4}).
Statistics of this lines (see figure~\ref{fig5}) agrees very well
with the corresponding curve of light intensity behind the circular
disc \cite{gondran:AJP:2010,rinard:AJP:1976} determined by taking
the square of the field function $\Psi(x,y,z)$, which is given by
the Rayleigh-Sommerfeld formula and Babinet's principle
\cite{gondran:AJP:2010,sommerfeld-bk},
\be
 \Psi(x,y,z) = \Psi_0 \left\{ e^{ikz}
  + \int_S \frac{e^{ikr}}{r}\ \! \left(1 - \frac{1}{ikr}\right)
    \cos \theta dx_M dy_M \right\} ,
 \label{eq18}
\ee
where $r = \sqrt{(x-x_M)^2 + (y-y_M)^2 + z^2}$, $\cos \theta = z/r$,
$k = 2\pi/r$ and integration is taken on the surface of the opaque
disc $S$.

Due to circular symmetry the integration over two variables in
(\ref{eq18})
may be reduced to the integration over one variable. This simplifies
and makes faster the numerical evaluation of the field function and
the trajectories. In addition, this simplified formula makes
possible the analysis of the dependence of the field function on the
longitudinal z-coordinate. The details of this study constitute the
subject of a forthcoming paper, where we will also present the
Bohmian trajectories corresponding to a recent Poisson-Arago spot
experiment performed with molecules \cite{holst:PRA:2009}.

%%%%%%%%%%%%%%%%%%%%%%%%%%%%%%%%%%%%%%%%%%%%%%%%%%%%%%%%%%%%%%%%%%%%%%%
%%%%%%%%%%%%%%%%%%%%%%%%%%%%%%%%%%%%%%%%%%%%%%%%%%%%%%%%%%%%%%%%%%%%%%%

\ack

MD, MB and DA acknowledge support from the Ministry of Science of
Serbia under Projects OI171005, OI171028 and III45016. ASS
acknowledges support from the Ministerio de Econom{\'\i}a y
Competitividad (Spain) under Projects FIS2010-22082 and
FIS2011-29596-C02-01, as well as for a ``Ram\'on y Cajal'' Research
Fellowship.

%%%%%%%%%%%%%%%%%%%%%%%%%%%%%%%%%%%%%%%%%%%%%%%%%%%%%%%%%%%%%%%%%%%%%%%
%%%%%%%%%%%%%%%%%%%%%%%%%%%%%%%%%%%%%%%%%%%%%%%%%%%%%%%%%%%%%%%%%%%%%%%

\section*{References}

%\bibliography{references}

\providecommand{\newblock}{}

\end{document}